\documentclass[jcp,aip,amsmath,amssymb,preprint]{revtex4-1}
\usepackage{graphicx}
\usepackage{epstopdf}

\begin{document}

\preprint{AIP/123-QED}

\title{Collapse transition of a square-lattice polymer with next nearest-neighbor interaction}

\author{Jae Hwan Lee}
\affiliation{School of Systems Biomedical Science and Department of Bioinformatics and Life Science, Soongsil University, Seoul 156-743, Korea}

\author{Seung-Yeon Kim}
\email{sykimm@cjnu.ac.kr}
\affiliation{School of Liberal Arts and Sciences, Chungju National University, Chungju 380-702, Korea}

\author{Julian Lee}
\email{jul@ssu.ac.kr}
\affiliation{School of Systems Biomedical Science and Department of Bioinformatics and Life Science, Soongsil University, Seoul 156-743, Korea}

\date{\today}

\begin{abstract}

We study the collapse transition of a polymer on a square lattice
with both nearest-neighbor and next nearest-neighbor interactions,
by calculating the exact partition function zeros up to chain length
36. The transition behavior is much more pronounced than that of the model with nearest-neighbor interactions only. The crossover exponent and the transition temperature are estimated from the scaling behavior of the first zeros with
increasing chain length. The results suggest that the model is of the same universality class as the usual $\theta$ point described by the model with only nearest-neighbor interaction.

\end{abstract}



\maketitle

\section{Introduction}

A flexible polymer chain in a dilute solution is influenced by both
hydrophobic interactions between the monomers and the excluded
volume effect. The attractive interactions are neglected at high
temperatures or in a good solvent, but become significant as the
temperature $T$ is lowered. As $T$ reaches a special temperature
$\theta$, the linear polymer undergoes an abrupt change from an
expanded conformation for $T>\theta$ to a fully compact conformation
for $T<\theta$~\cite{F67,CD91,dG75}. Long polymer in a good solvent
is a critical system, and the collapse transition at $T=\theta$ has
been identified as a tricritical transition~\cite{G73,dG75}. The
$\theta$ point behavior is well-described by self-avoiding walks
 with attractive interaction energy assigned for each pair of
nonbonded nearest-neighbor  monomers. The tricritical exponents take
the mean-field values for $d>3$, and there are logarithmic
corrections at $d=3$~\cite{dG75,G73,S75,B82,S86,DS87}. A great deal
of studies have been performed to understand the nature of the
collapse transition in two
dimensions~\cite{S75,B82,KF84,BBE85,DS85,S86,P86,CJMS87,DS87,PCJS88,DS88,SSV88,DS88b,SS88,CD89,ML89,DS89b,
PCJS89,DS89,VSS91,GH95,BBG98,NKMR01,ZOZ08,CDC09,GV09,L04,LKL10,CGPP11},
which is expected to exhibit much more non-trivial behavior than its
higher dimensional counterparts.

In this work, we study the collapse transition of a polymer on a
square lattice, with both nearest-neighbor (NN) and next
nearest-neighbor (NNN) interactions present, by calculating the
exact partition functions up to chain length $N=36$. We estimate the
crossover exponent and the transition temperature from the zeros of
the partition function, and also from the specific heat. Although
the  method of partition function zeros became one of the most popular
tools for studying the critical phenomena with the advancement of
computational power~\cite{BDL05,CC05}, there are few works where
partition function zeros of lattice polymers were calculated. For
examples, exact partition function zeros were computed for the
simple-cubic lattice up to chain length 13~\cite{FJB75}, for the
face-centered lattice up to chain length 9~\cite{Rap}, and for the
square lattice up to chain length 36~\cite{L04,LKL10}. Only NN
interactions were present in these works.  In fact, the current work
is the first instance where a {\it square}-lattice polymer with NNN
interactions is ever studied. It was only on a hexagonal lattice
that models with NNN interactions were studied previously
~\cite{DS87,PCJS88,DS88,PCJS89,DS89,VSS91}.

By introducing NNN interactions, the transition behavior is much more pronounced than that of the model only with NN interactions~\cite{LKL10}.
The results suggest that the model belongs to the same universality
class as the one described by the model with only NN interactions.

\section{The Number of Conformations}

Conformations of a polymer chain with $N$ monomers are modeled
as a two-dimensional self-avoiding chain of length $N$ on a square lattice.
The position of the monomer $i$ is given by ${\bf r}_i = (k,l)$,
where integers $k$ and $l$ are the Cartesian coordinates relative to an arbitrary origin.
Chain connectivity requires $\vert {\bf r}_i-{\bf r}_{i+1} \vert = 1$, i.e., bond length is unity.
Due to the excluded volume, there can be no more than one monomer on each lattice site,
${\bf r}_i \neq {\bf r}_j$ for $i \neq j$.
The attractive hydrophobic interaction is incorporated
by assigning the energies $-\epsilon_1 < 0$ and $-\epsilon_2 < 0$ for each non-bonded NN and NNN contact between monomers.
The resulting Hamiltonian is
\begin{equation}
    {\cal H} = -\epsilon_1 \sum_{i<j} \Delta ({\bf r}_i, {\bf r}_j) - \epsilon_2 \sum_{i<j} \tilde{\Delta} ({\bf r}_i, {\bf r}_j),
\label{hamiltonian}
\end{equation}
where
\begin{equation}
\Delta ({\bf r}_i, {\bf r}_j) = \left\{
    \begin{array}{ll}
        1 & ~~{\rm if} ~~ |i-j| > 1 ~~ {\rm and} ~~ |{\bf r}_i-{\bf r}_j| =1,\\
        0 & ~~{\rm otherwise},
    \end{array} \right.
\end{equation}
and
\begin{equation}
\tilde{\Delta} ({\bf r}_i, {\bf r}_j) = \left\{
    \begin{array}{ll}
        1 & ~~{\rm if} ~~ |{\bf r}_i-{\bf r}_j| = \sqrt{2},\\
        0 & ~~{\rm otherwise}.
    \end{array} \right.
\end{equation}
The result when only NN interactions are present, corresponding to the $\theta$ point~\cite{LKL10}, can be reproduced by
putting $\epsilon_2 = 0$.
We consider the case with $\epsilon_1 = \epsilon_2 \equiv \epsilon$.
The energy of the system is then $E=-\epsilon (K_1 + K_2 ) \equiv -\epsilon K$,
where $K_1$ and $K_2$ are the number of contacts between NN and NNN monomers, respectively.

Here we define the reduced number of conformations $\omega_N(K)$,
where conformations related by rigid rotations, reflections, and translations are regarded as equivalent,
and counted only once.
On the other hand, due to an assumption that the polymer chain has an intrinsic direction,
the conformations with reverse labels $i \leftrightarrow N-i+1$ for all $(i=1,2, \cdots, N)$ are considered distinct.
It is easy to see that
the total number of conformations generated by rotations and reflections
from a given conformation is eight, except for the straight chain where
the total number of conformations generated by rotations and reflections is four due to invariance with respect to reflection perpendicular to the chain.
The total number of conformations $\Omega_N(K)$ is obtained from $\omega_N(K)$ as follows:
\begin{equation}
\Omega_N (K) = \left\{
    \begin{array}{ll}
        8\omega_N (K) -4    & \quad {\rm if} ~~ K = 0,\\
        8\omega_N (K)       & \quad {\rm otherwise}.
    \end{array} \right.
\label{total}
\end{equation}
Thus, one can achieve about eight-fold reduction in the computing time by enumerating the reduced number of conformations $\omega_N(K)$ instead of $\Omega_N(K)$~\cite{LKL10}.
We obtained $\omega_N (K)$ up to $N=36$
by the help of a parallel algorithm
classifying conformations by sizes of rectangles they span~\cite{LKL11}.

\section{partition function zeros in the complex temperature plane}

Yang and Lee~\cite{YL52} first introduced the concept of the partition function zeros
 in the complex fugacity plane,
and found a mechanism for the occurrence of phase transitions in thermodynamic limit.
Later, Fisher~\cite{F65} showed that
the partition function zeros in the complex temperature plane are very important
in understanding phase transitions.
For system exhibiting the temperature-driven phase transition,
the locus of Fisher zeros forms a line
and crosses the positive real axis in thermodynamic limit.
The intersection point of the locus with the positive real axis corresponds to the critical temperature.
The zeros closest to the positive real axis are called the \textit{first} zeros,
which approach the positive real axis as the system size increases.

The partition function of our model is
\begin{equation}
    Z =\sum  e^{-\beta \cal H} = \sum_{K}  \Omega_N(K) y^{K}, \label{PF}
\end{equation}
where  $y \equiv \exp(\beta \epsilon)$ and $\beta \equiv 1/k_B T$.
We see that since $K$ is bounded, the partition function (\ref{PF}) is a $n$-th order polynomial of $y$ where $n$ is the maximum value of $K$. The partition function zeros $y_i~ (i=1,2,\cdots, n)$ are then obtained by solving the polynomial equation $Z(y) =0$. The solution was found with \textsc{mathematica}.
As can be seen from Fig.~\ref{first_zeros},
the first zeros approach the positive real axis in the complex temperature plane as polymer length increases.

\section{The Scaling Behavior and the Critical Exponent}

\begin{figure}
\includegraphics*[width=.9\textwidth]{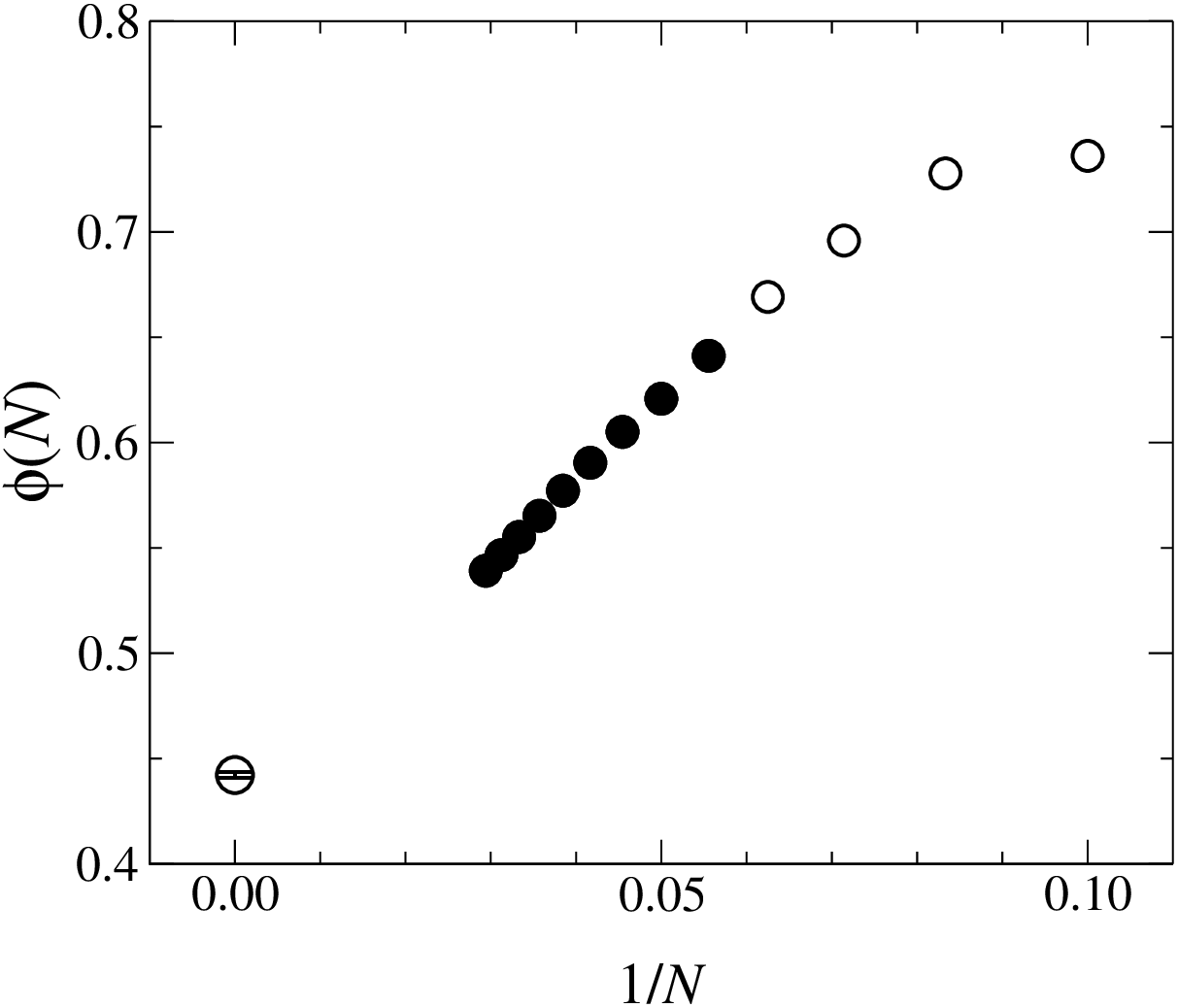}
\caption{The finite size approximations of the crossover exponent,
$\phi(N)$, are shown as a function of $1/N$ for even $N$ with $10
\le N < 18$ (open circles) and $N \ge 18$ (solid circles). The value
of $\phi=0.4422(14)$ for $N \to \infty$ (the open circle with an
error bar) is estimated by the BST extrapolation for $N \ge 18$.}
\label{nphi}
\end{figure}

Near the critical temperature $T_c$,
the radius of gyration (or the end-to-end distance) $R_N$ of a polymer chain with $N$ monomers
is generally expressed by the scaling theory~\cite{dG75,S75},
\begin{equation}
    \langle R_N^2 \rangle \sim N^{2\nu} f(\tau N^{\phi}),
\label{scaling}
\end{equation}
where the reduced temperature is defined as $\tau \equiv \left \vert T-T_c \right \vert / T_c$
and the scaling function $f(x)$ behaves as follows:
\begin{eqnarray}
    f(x) &=& \left\{
    \begin{array}{lll}
        x^{(6/(d+2)-2\nu)/\phi} & \quad &   \mathrm{if}~~x \to \infty,  \\
        \mathrm{const.}             & &         \mathrm{if}~~x \to 0,       \\
        x^{(2/d - 2\nu)/\phi}       & &         \mathrm{if}~~x \to -\infty.

    \end{array}
        \right.
\end{eqnarray}
The exponent $\nu$ represents the geometrical properties of a polymer,
and the crossover exponent $\phi$ describes how rapidly the system undergoes the transition
as $T$ approaches $T_c$.
The crossover exponent $\phi$ also describes how rapidly the first zeros approach the positive real axis as $N$ increases~\cite{LKL10},
\begin{equation}
    \mathrm{Im}[y_1(N)] \sim N^{-\phi},
\end{equation}
where $y_1(N)$ is a first zero for a polymer chain with $N$
monomers. In finite-size systems with even $N$, the crossover
exponent is approximated as
\begin{equation}
    \phi(N) = - \frac{\ln\{{\rm Im}[y_1(N+2)]/{\rm Im}[y_1(N)]\}}{\ln\{(N+2)/N\}},
    \label{phi}
\end{equation}
which reduces to the exact value of $\phi$ in $N \to \infty$ limit,
estimated by using the Bulirsch-Stoer (BST)
extrapolation~\cite{BS64}. We obtain 0.4422(14) for the
crossover exponent as shown in Fig.~\ref{nphi}, where the estimated
error could be further reduced by removing unreliable data obtained
from $N<18$.  The error is estimated by examining the
robustness of the extrapolated value with respect to perturbations
of the data points, but it is not a statistically rigorous
confidence level~\cite{BS64,LKL10}. Therefore, we estimated the error 미내 by slightly changing the ratio of  NNN and NN interactions, $R \equiv \epsilon_2/\epsilon_1$, which we set to 1 in the current work. We change $R$ by 0.5, and get $\phi=0.428$ for both $R=0.5$ and $1.5$.  If we assume that $R$ is irrelevant and combine the results for $R=0.5$, $1.0$, and $1.5$, the resulting range of the crossover exponent is $0.428 \le \phi \le 0.442$.  The result is consistent with the conjectured exact value of $\phi=3/7=0.4286$ obtained from hexagonal lattice with random annealed forbidden faces\cite{DS87}, as well as our previous estimate from the model with $NN$ interactions only, $\phi=0.422(12)$, suggesting that they belong to the same universality class.  More extensive analysis for various values of $R$ is postponed for a future study.

Without additional information, we assumed the leading
finite size correction to $\phi$ is of order $O(N^{-1})$ when
performing the BST procedure. We estimated the range of $\phi$ also by changing the leading exponent of the extraopolating function. With $R=1$ fixed, we performed BST extrapolation with the leading finite size correction of order    $O(N^{-\omega})$ with $\omega=0.5$ and $1.5$. We get $\phi=0.418$ and $0.458$ for $\omega=0.5$ and $1.5$ respectively, and combining these results with that for $\omega=1.0$, we get $0.418 \le \phi \le 0.458$, again consistent with both the conjectured exact value and the estimate from the model with $NN$ interactions only. 
Again, there is no evidence
that our model belongs to a universality class different from that of the model with NN interactions only.

\begin{figure}
\includegraphics*[width=.9\textwidth]{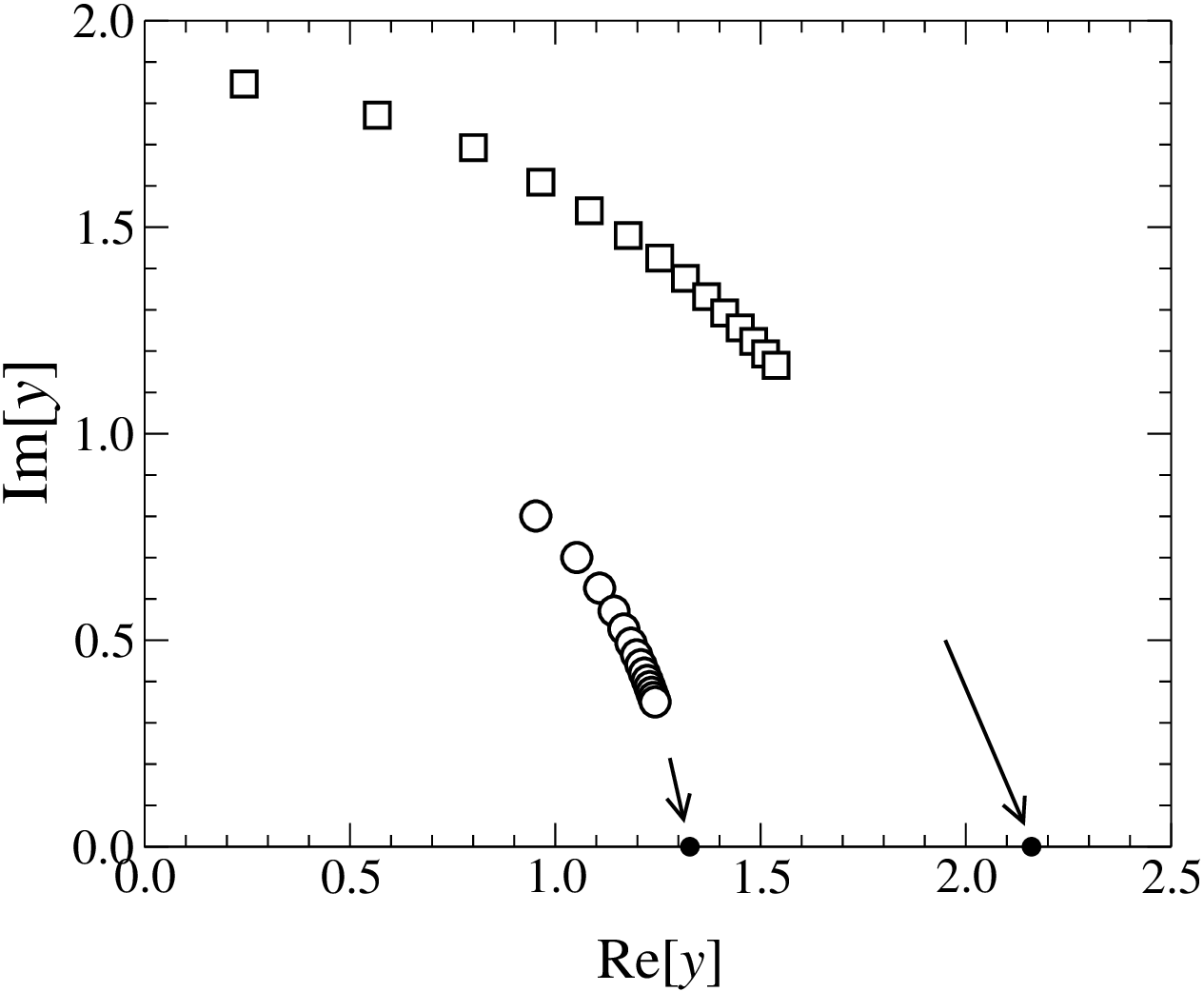}
\caption{Positions of the first zeros in the first quadrant of the
complex temperature ($y = e^{\beta \epsilon}$) plane for $N =
10,~12,\cdots,~36$. Open circles indicate the results when both NN
and NNN interactions are present, and open squares are those for the
model with NN interactions only. Two dots indicated by arrows are
the corresponding values of $y_c$.} \label{first_zeros}
\end{figure}

\begin{figure}
\includegraphics*[width=.9\textwidth]{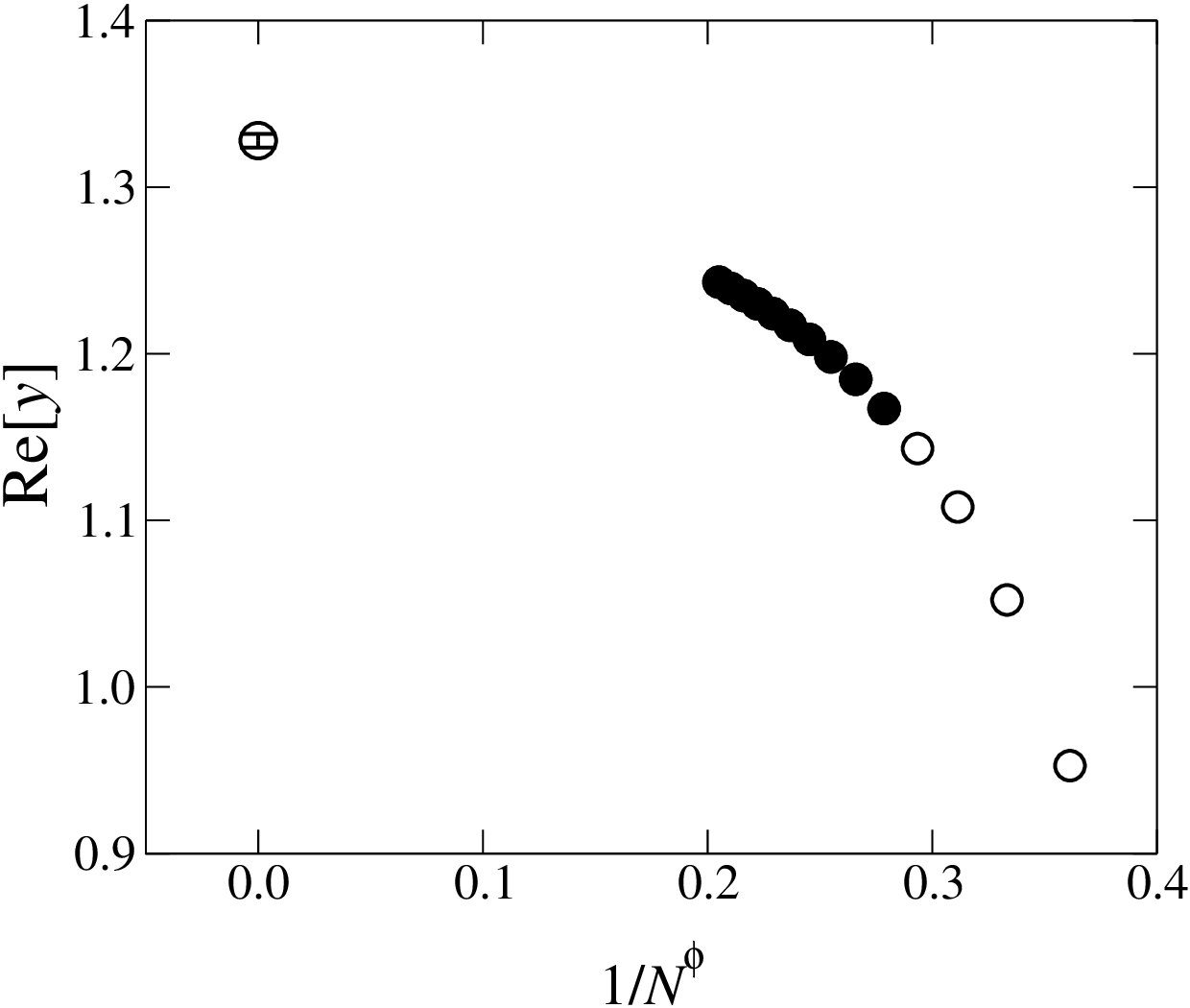}
\caption{The real parts of the first zeros are shown as a function
of $1/N^{\phi}$ for even $N$ with $10 \le N < 18$ (open circles) and
$N \ge 18$ (solid circles). The value of $y_c = 1.3279(41)$ (the
open circle with an error bar) for $N \to \infty$ is estimated by
the BST extrapolation for $N \ge 18$ with $\phi=0.4422$.}
\label{nre}
\end{figure}

The real parts of the first zeros can be used to estimate the
critical temperature $y_c$, by estimating the point they approach in
the limit of $N \to \infty$,
\begin{equation}
    \mathrm{Re}[y_1(N)]-y_c \sim N^{-\phi},
\label{yc}
\end{equation}
with the value of $\phi$ obtained above. The value of $y_c$,
obtained by extrapolating the data for even $N$ with $N \ge 18$,  is
$1.3279(41)$,  which corresponds to $T_c /\epsilon = 3.526(39)$
(Fig.~\ref{nre}). It is also shown in Fig.~\ref{first_zeros} along
with the result for the model where only NN interactions are
present~\cite{LKL10}, corresponding to $y_c = 2.16(18)~(T_c
=1.30(17))$. The transition temperature becomes much higher when
additional attractive NNN interactions are included, which is to be
expected. We obtain $y_c = 1.3288(41)$ with the conjectured exact
value $\phi = 3/7$~\cite{DS87}, which is not much different from the
result above. As can be seen from Fig.~\ref{first_zeros}, the
transition behavior is much more visible when we introduce NNN
interactions.

\section{specific heat}

\begin{figure}
\includegraphics*[width=.9\textwidth]{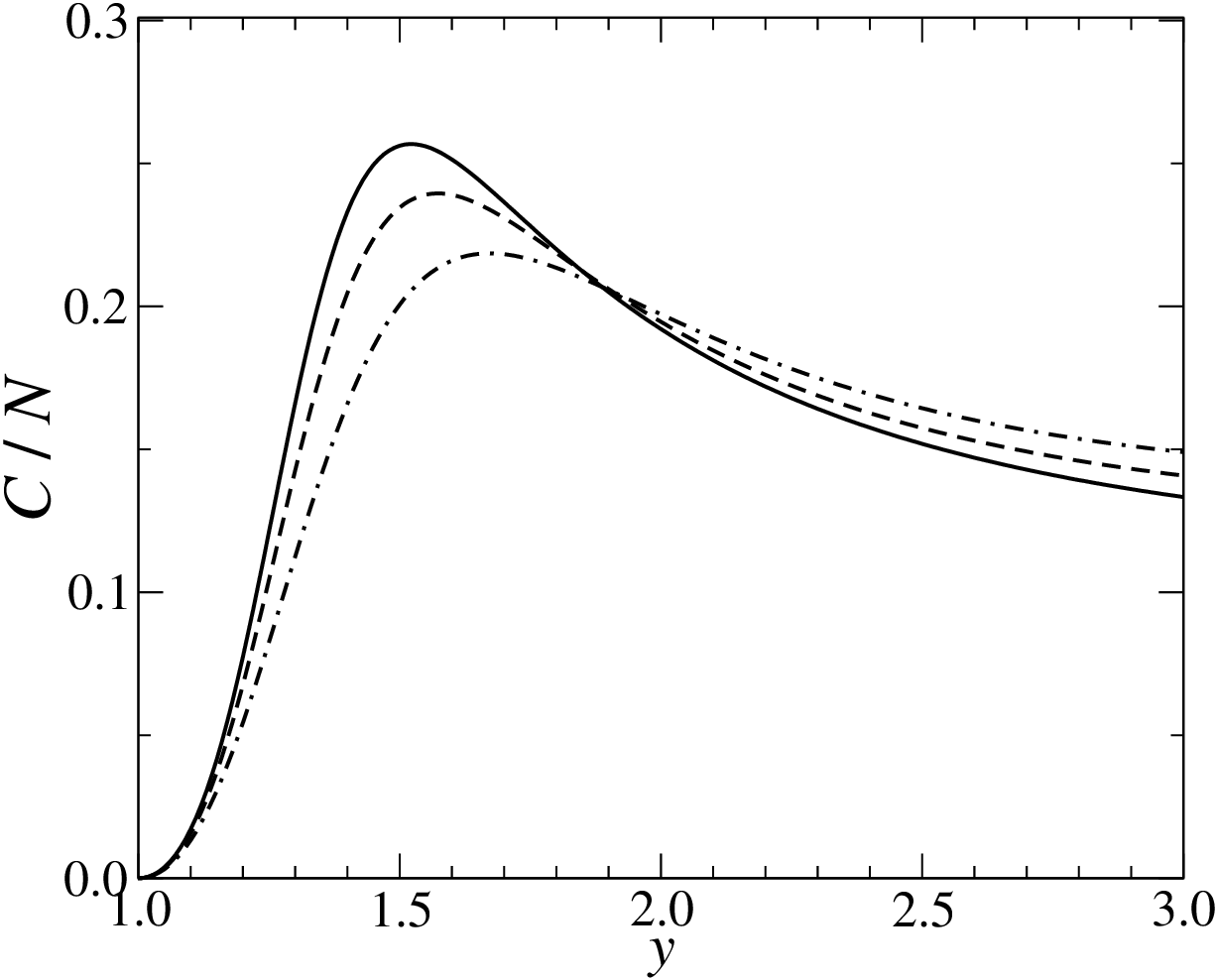}
\caption{The specific heat for $N=20, 28$, and 36 from bottom to top.}
\label{cy}
\end{figure}

\begin{figure}
\includegraphics*[width=.9\textwidth]{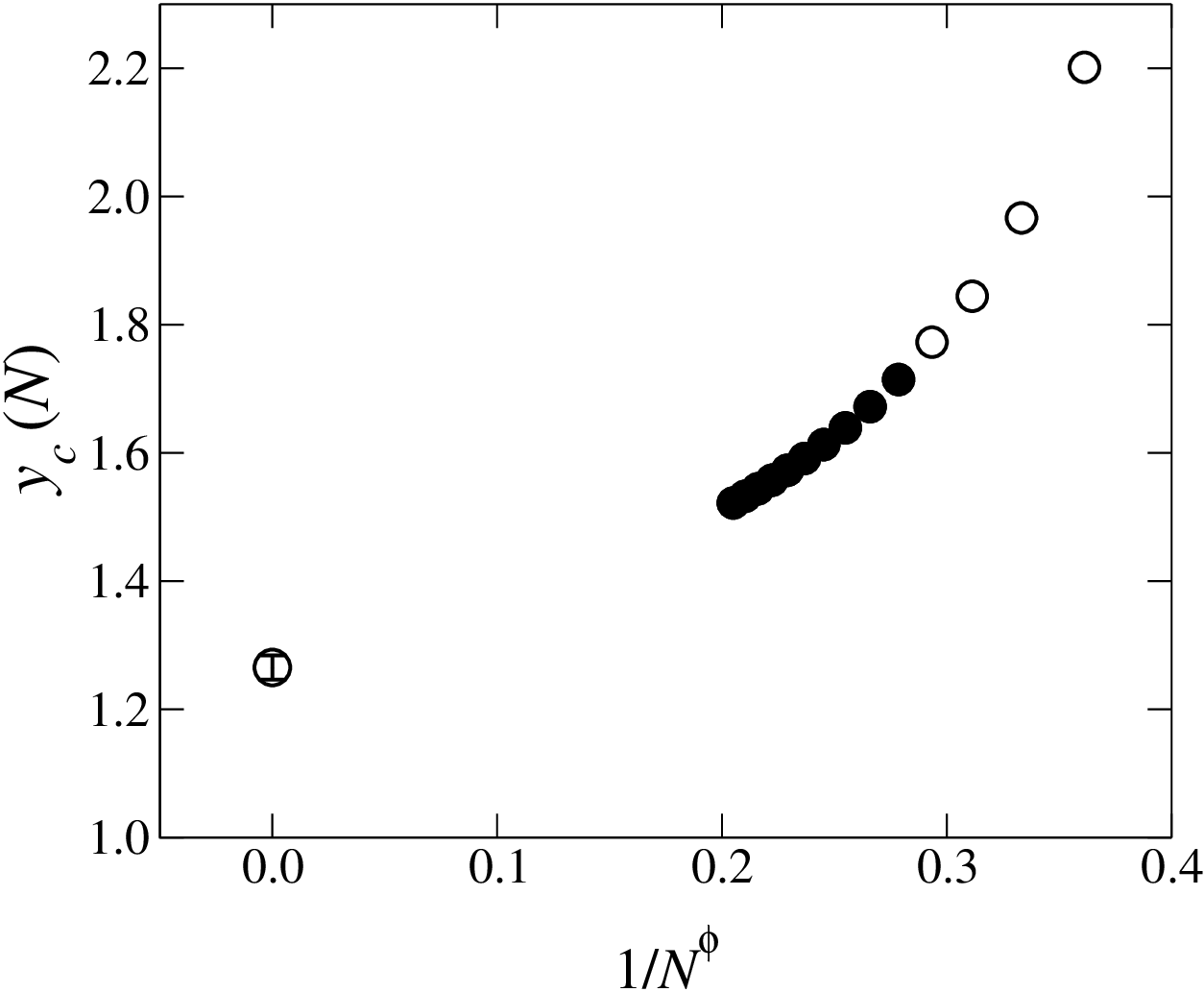}
\caption{The finite size approximation of $y_c$ obtained from the
specific heat, $y_c (N)$, are shown as a function of $1/N^{\phi}$
for even $N$ with $10 \le N < 18$ (open circles) and $N \ge 18$
(solid circles). The value of $y_c (\infty) = 1.265(19)$ (the open
circle with an error bar) is estimated by the BST extrapolation for
$N \ge 18$ with $\phi=0.4422$.} \label{yn}
\end{figure}

Now we estimate the critical temperature $y_c$ again
by analyzing the behavior of the specific heat per monomer,
for comparison with the result obtained from the partition function zeros. The specific heat per monomer is
\begin{eqnarray}
    \frac{C(T,N)}{\epsilon^2 N} &=& \frac{1}{\epsilon^2 N}\frac{\partial E}{\partial T} \nonumber \\
        &=& \frac{\beta^2}{\epsilon^2 N} \frac{\partial^2 \ln Z}{\partial \beta^2} \\
        &=& \frac{(\ln y)^2}{N}\left[ \frac{\sum_K K^2 \Omega_N (K) y^K}{\sum_K \Omega_N (K) y^K}-\left(\frac{\sum_K K \Omega_N (K) y^K}{\sum_K \Omega_N (K) y^K}\right)^2\right],  \nonumber
\end{eqnarray}
which is plotted in Fig.~\ref{cy} as a function of $y$ for several values of $N$.
The finite $N$ approximation of the transition point, $y_c(N)$, is obtained from the condition
$\frac{\partial C}{\partial y} = 0$.
We observe a peak around $y \simeq 1.5$, which becomes sharper as $N$ increases.
By applying the BST extrapolation  to the finite-size scaling
\begin{equation}
    y_c(N) - y_c(\infty) \sim N^{-\phi},
\end{equation}
we obtain the transition point $y_c(\infty) = 1.265(19)$,
equivalent to $T_c/\epsilon = 4.25(29)$,
where the data for even $N$ with $18 \le N \le 36$ were used.
$y_c(N)$ is displayed in Fig.~\ref{yn} as a function of $1/N^\phi$,
along with the extrapolated value $y_c(\infty)$.
The current result is not drastically different from that obtained by the partition function zeros,
but the precision is lower due to the fact that
the specific heat is riddled by noisy contributions from zeros other than the first ones~\cite{LKL10}.

\section{Discussions}

In this work, we studied the collapse transition of a square-lattice
polymer with both NN and NNN interactions, by calculating the exact
partition function zeros up to chain length $N=36$. The crossover
exponent $\phi$ and the transition temperature $T_c$ were obtained
by examining their scaling behavior with increasing chain length. We
estimated $T_c$ also by calculating the specific heat from the exact
partition function. Our results suggest that the polymer with
both NN and NNN interactions on a square lattice belongs to the
$\theta$ universality class described by the model where only NN
interactions are present, but by introducing NNN interactions, the
transition behavior becomes more pronounced than the model with only NN
interactions~\cite{LKL10}.

\begin{acknowledgments}
This work was supported by Mid-career Researcher Program through NRF grant funded by the MEST (No.2010-0000220).
\end{acknowledgments}


%
%
\newpage
\newpage
\newpage
\newpage
\newpage
\end{document}